\documentclass[prl,twocolumn,epsf]{revtex4}
\usepackage{graphicx}% Include figure files
\usepackage{dcolumn}% Align table columns on decimal point
\usepackage{bm}% bold math
\usepackage{hyperref}% Auto-generate hyperlinks

\newcommand{\beq}{\begin{eqnarray}}
\newcommand{\eeq}{\end{eqnarray}}
\begin{document}
\noindent
{\bf Doublon-Holon Bound States in a Half-filled Band}

In a recent Letter, Sakai, et al.\cite{sakai} performed a numerical
study of the Hubbard model on a square lattice in attempt to command a
clear view of a doped Mott insulator.  Among
the results they present as being new is the `` 
non-trivial finding''  that the spectral weight immediately above the
chemical potential in the lower
Hubbard band, $\Lambda(x)$, increases rapidly with hole doping ($x$) at the expense of spectral
weight in the upper Hubbard band.  Although this result is presented
without reference to any prior work, it is far from new.  It was
 first observed in a cuprate by  Chen, et
 al.\cite{chen} in 1991 and was immediately explained by Eskes, et
 al.\cite{sawatzky} as a ubiquitous phenomenon in a doped Mott insulator.  In fact, the density
 of states in Fig. (3a) of Sakai, et al.\cite{sakai} is identical to
 that in Fig. 2 in the second of Ref. 3. 

Sakai, et al.\cite{sakai} also
propose that $\Lambda (x)$ is
described by $2x+2n_d$ (counting both spins) where $n_d$ is the number of doubly occupied
sites. While the dynamical part of $\Lambda(x)$ is
certainly mediated by double occupancy, it is {\it not} equivalent to it as can be seen from the perturbative
expression, 
\beq
\Lambda(x)=2x-\frac{2t}{U N}\sum_{\langle ij\rangle,\sigma} \langle
c_{i\sigma}^\dagger c_{j\sigma}\rangle,
\eeq
 derived by Harris and Lange\cite{harris} in 1967 (also not cited).
 This expression contains a static part ($2x$) because each hole creates
two addition states at low energy and
a dynamical part which is strictly positive and proportional to the
expectation value of the kinetic energy. In fact, all orders of
perturbation theory\cite{sawatzky,harris} increase $\Lambda(x)$ beyond its atomic limit of $2x$.    The result of Sakai, et
al.\cite{sakai} follows from the perturbative analysis only if the virial theorem holds for the
Hubbard model, which it does not.  As a
consequence, any relationship between the dynamical part of $\Lambda(x)$ and double
occupancy is not one of equivalence but rather supervenience. 

In fact, Sakai, et al.\cite{sakai} are well aware that their analysis
is problematic.  They note correctly, that on their account,
$\Lambda(x=0)=2n_d$ at half-filling, which is non-zero rather than
vanishing as would be required by the Mott gap. To
correct this flaw in their interpretation of $\Lambda(x)$, they state, without any kind of proof, that some
type of bound state between doublons and holes saves the day by
transferring spectral weight back to the upper Hubbard band. It is
unclear what this statement means since there is no energy scale associated
with $n_d$.  It is simply a number.  Further, Sakai, et al.\cite{sakai}
offer no explanation as to why their mechanism for the pseudogap, screening-induced
degradation of the binding energy,  would lead to an abrupt rather
than the more physically reasonable gradual reduction of the Mott gap.

However, the claim that bound states between doublons and holons
mediate the Mott gap is not without merit because without them, the
holons (arising entirely from quantum fluctuations) would be mobile thereby destroying the Mott gap and
preventing antiferromagnetic order.  As a consequence, establishing that such bound
states are the propagating degrees of freedom in a half-filled band is
the essence of the Mott problem, provided, of course, that the band
structure of these elementary excitations yields the gapped spectrum
indicative of Mottness.  To this end, the Wilsonian program of
integrating out the high energy scale
indispensable and has been recently carried out exactly\cite{lowen1}
 for a half-filled band described by the
Hubbard model.  The exact
low-energy Lagrangian demonstrates\cite{lowen1} that while there are no bare degrees of
freedom which propagate, composite excitations corresponding to
bound doublon/holon or double hole/electron pairs
do propagate.  The spectral
weight\cite{lowen1} of the doublon/holon pairs turns on for 
$U/2-4t<\omega<U/2+4t$ while that for the double-hole/electron pair is non-zero 
for $-U/2-4t<\omega<-U/2+4t$, thereby describing the dispersions for the
upper and lower Hubbard bands, respectively.  For $U<8t$ ($8t$ the
bandwidth) the gap for the composite excitations (not that for the electrons) vanishes giving rise
to a discontinuous jump in the density of states at the chemical
potential to the corresponding value in the non-interacting system.  At
finite doping, such bound states still persist and generate a
pseudogap\cite{lowen1}, though with a new energy scale not related to
the Mott gap via the screening mechanism envisioned by Sakai, et al.\cite{sakai}.

\vspace{12pt}
\noindent Philip Phillips\\
Loomis Laboratory of Physics\\
University of Illinois\\
Urbana, Il. 61801-3080

\acknowledgements I acknowledge the NSF DMR-0605769 for partial support.

\end{document}